\documentclass[12pt]{article}
\usepackage{epsfig}

\parskip=\medskipamount % space between paragraphs (TeX)
plus.3em minus.5em \abovedisplayshortskip=.5em plus.2em minus.4em
\renewcommand{\thesection}{\arabic{section}}

\catcode`@=11

\@addtoreset{equation}{section} \@addtoreset{equation}{subsection}
\def\theequation{\ifnum\value{section}=0 \arabic{equation}\ignorespaces
\else \ifnum\value{section}=-1 A.\arabic{equation}\ignorespaces
\else \ifnum\value{subsection}=0
\thesection.\arabic{equation}\ignorespaces \else
\thesection.\arabic{subsection}.\arabic{equation}\ignorespaces
                             \fi
                        \fi
                   \fi}

{\catcode`\'=\active \def'{{}^\bgroup\prim@s}}

\catcode`@=12

\newcommand{\bq}{\begin{equation}}
\newcommand{\be}{\begin{equation}}
\newcommand{\fq}{\end{equation}}
\newcommand{\ee}{\end{equation}}
\newcommand{\bqr}{\begin{eqnarray}}
\newcommand{\beqs}{\begin{eqnarray}}
\newcommand{\fqr}{\end{eqnarray}}
\newcommand{\eeqs}{\end{eqnarray}}

\newcommand{\rf}[1]{(\ref{#1})}

\def\bop#1{\setbox0=\hbox{$#1M$}\mkern1.5mu
    \vbox{\hrule height0pt depth.04\ht0
    \hbox{\vrule width.04\ht0 height.9\ht0 \kern.9\ht0
    \vrule width.04\ht0}\hrule height.04\ht0}\mkern1.5mu}
\begin{document}
\thispagestyle{empty}

\begin{flushright}
\begin{tabular}{l}
% TEP- \\
hep-th/0503200 \\
\end{tabular}
\end{flushright}

\vskip .6in
\begin{center}

{\bf  Integer and Rational Solutions to Polynomial Equations}

\vskip .6in

{\bf Gordon Chalmers}
\\[5mm]
% {\em address \\
%      address \\
% Los Angeles, CA } \\

{e-mail: gordon@quartz.shango.com}

\vskip .5in minus .2in

{\bf Abstract}

\end{center}

A formalism is given to count integer and rational solutions to
polynomial equations with rational coefficients.  These polynomials
$P(x)$ are parameterized by three integers, labeling an elliptic curve.
The counting of the rational solutions to $y^2=P(x)$ is facilitated by
another elliptic curve with integral coefficients.  The problem of 
counting is described by two elliptic curves and a map between them.

\newpage
\setcounter{footnote}{0}

The integer and rational solutions to polynomial equations with rational
coefficients is relevant to many areas of physics, mathematics, and
applied science.  The counting of these solutions seems problematic
without direct solving of the equations.  One conjecture relates the
expansion of certain Hecke modular forms (L-series) around unity to the
cardinality of the rational solutions to $P(x)=y^2$.  A direct counting
of the solutions, and their characterizations is more interesting and
relevant.

In this work, all rational polynomial equations are examined in one
context.  The counting of solutions $p/q$ for a given $q$ to the equation
$P(x)=m/n$ is formulated in this work.  Several functions $f_i$ are
required to analyze the complete set of polynomial equations.
The symmetry and uniformizations of the counting problem is described by
the general functional form of these functions, which consist of a map
from one elliptic curve to another.  The maps are not given here; however,
a determination would certainly be of interest for both practical
applications of the counting and for formal extensions 
of algebra \cite{apps}.

As an example of the counting problem, consider the expansion of the free
energy of a certain statistical model as described in \cite{KnotInv} (with 
related work in \cite{KnotPaper}),

\bqr
\sum_{x=1}^M \exp{(-\sum_{i=1}^N b_i x^i \tau)} =
 \sum_p \Delta(p) e^{-p\tau} \ .
\label{onevariable}
\fqr
The statistical free energy in \rf{onevariable} involves a one parameter
sum over integers $x$, which admit a $\tau=1/k_b T$ expansion at small
temperature.  The expansion is facilitated by the integer solutions
to

\bqr
\sum_{i=1}^N b_i x^i = P(m)=p \ ,
\label{zeroset}
\fqr
and a counting of the number of solutions $m$.  The formalism of the
model in \rf{onevariable} specifying the $b_i$ coefficients (which are
in general rational) gaurantees that for integer $x$ the value of $P(x)$
is an integer.  The number of solutions at level $p$ to the equation in
\rf{zeroset} defines the function $\Delta(p)$.  The function $\Delta(p)$
is defined by its values at $M$ points, and hence can also be characterized
by a polynomial of degree $M$.  These polynomials in \rf{zeroset}
together with their solutions define quasi-modular and modular forms
via the expansion in \rf{onevariable}.

The example describes one purpose of the counting of the divisors of the
rational polynomials, and how the countings of $m$ to $P(m)=p$ are
described by a polynomial $\Delta(p)$.  The polynomial aspect to $\Delta(p)$
is examined in further detail here, albeit in a different context and
generalized to the set of all polynomials with rational coefficients.

\noindent {\it Counting of Solutions}

Algebraic equations $P(x)=y^2$ are examined, with the degree of $P(x)$
and integer $n$.  Rational solutions of the form $x=s/t$ and $y^2=p/q$
are considered.  In general $y^2$ may be taken to $y^d$, generating a
further restriction on the $y$ values.

The general polynomial equation

\bqr
P(x) = a_n x^n + a_{n-1} x^{n-1} + \ldots + a_0 \ ,
\fqr
may be specified by three integers, in the case of rational coefficients
$a_i = p_i/q_i$.  For example, first specify a base $m$ for $B$.  Then
specify two polynomials parameterizing the numerators and denominators
of the coefficients pertaining to $P(x)$ via,

\bqr
N=\prod_{j=1}^n (x-p_i) \qquad D=\prod_{j=1}^n (x-q_i)  \ .
\fqr
The base $x$ must be chosen larger than any of the coefficients
$\prod_i p_i$ and $\prod_i q_i$.  The expansion of the numbers $M$ and
$N$ in the base $x$ follows from,

\bqr
N = \sum_{k=1}^n N_k x^k   \qquad  D = \sum_{k=1}^n D_k x^k \ ,
\fqr
with $N_k$ and $D_k$ smaller than the integer $x$.  In this base, the
numbers $N$ and $D$ uniquely parameterized the coefficients $N_k$ and
$D_k$.

These three numbers $B$, $N$, and $D$ parameterize the polynomial $P(x)$.
One may group these numbers into an algebraic elliptic curve $E$,

\bqr 
x^3 + Bx^2 + N x + D = y^2 \ , 
\fqr 
or  
\bqr
B x^2 + N x + D = y^2  \ ,
\label{elliptic}
\fqr
or into a 3-tuple $(B,N,D)$.  The elliptic curve is parameterized up to
relabeling of the three numbers.

As a comment, the parameterization of the rational polynomials is
not unique.  Another decomposition is to write the polynomial in terms
of a product and an integer, in base $B$: $a\prod (x-c_j) + M=N+M$, which
requires three numbers $(x,N,M)$.  The one used here seems natural, and
the former is related transcendentally.

To illustrate the procedure a counting map is generated for integer
solutions to $P(x)=p/q$.  The question to be examined is how many
rational solutions $x=s/t$, i.e. divisors, there are to this equation.

First rationalize the polynomial equation by multiplying the equation
by the denominators of the coefficients $t^n q \prod^n_{i=1} b_i$.
This generates a polynomial equation in $s$, $P_t(s)=p$, containing
integer coefficients.

The question to be examined is how many integer solutions there are
to this equation $P_t(s)=p$.  Label the count as $C_t^{p,q}(s)$, valid in
a range of values $s=q_1$ to $s=q_2$.  This count $C_t(s)^{p,q;q_1 q_2}$ is 
specified by another polynomial,

\bqr
Q_t(x) = b_{L} x^L + b_{L-1} x^{L-1} + \ldots b_0 = 0 \ ,
\label{countpoly}
\fqr
with $x$ ranging from $q_1$ to $q_2$ for example.  In principle, the
values $q_1$ and $q_2$ may be taken to $-\infty$ and $\infty$, as a
limit of the two integers; this results in infinite degree (or an
analytic function, possessing a Taylor series expansion with the
coefficients $b_j$.

By definition, the polynomial in \rf{countpoly} counts the solutions
$x=s/t$ to the polynomial equation $P(x)=p/q$, denoted by $C_t(s)^{p,q;q_1 q_2}$
in the range $s=q_1$ to $s=q_2$.  This polynomial $Q(x)$ is also specified
by an associated elliptic curve $E_C$, or a $3$-tuple,

\bqr
B_C x^2 + N_C x + D_C = 0 \ .
\label{rationalcount}
\fqr
The three integers $B_C$, $N_C$, and $D_C$ label the curve.

Given the elliptic curve $E_C$, the polynomial $Q(x)$ may be reconstructed
by expanding the numbers $N_C$ and $D_C$ in the base $B_C$.  In this manner
the elliptic curve counts the solutions $C_t^{p,q}(s)$ in the range $q_1$ to $q_2$.

The map required is a transformation of the two curves from

\bqr
B x^2 + N x + D = y^2  \quad \rightarrow \quad
 B_C x^2 + N_C x + D_C = y^2 \ ,
\label{elliptictrans}
\fqr
or $x^3 + B_i x^2 + N_i x + D_i =y^2$ (as with all of the other three tuple forms 
of the curves previously described).
This is an SL(2,Z)$_w$ transformation parameterized for example by three
numbers, or a complex rational number $f_{r/g}+ i f_{h/g}$.  Denote the map
by $M_t^{p,q;q_1q_2}$.

All of the information of the rational count $x=s/t$ to the polynomial
equation $P(x)=p/q$ is encoded in the map between the elliptic curves.
These are essentially fibrations of the polynomial equations, e.g. the
elliptic curves over the Riemann surfaces $P(x)=y^2$ or $P(x)=y^d$.  The 
question is then to do determine $(B_C,N_C,D_C)$ from $(B,N,C)$
specified by $t$ and $q_1,q_2$, and $p$ and $q$.

The simplest limit is $q_1=-\infty$, $q_2=\infty$, and $p=q=0$.  In this
case, there is a one parameter map $M_t$ labeled by the denominator of the
solution $x=s/t$, and the function $C_t(s)$ counts the $x$-solutions.
The most general mapping contains the five parameters, possibly a
function of the curve parameters $B,N,D$.

An interesting question is when the transformation from $E$ to $E_C$ depends
on the parameters $B,N,D$, and how it depends on these transformations.
Obviously there are relations between rational solutions $x=s/t$ to
polynomials with rational coefficients $P(x)=p/q$ and integer solutions
$x=m$ to polynomials $P_Z(x)=p$ with integer coefficients.  The various
mappings should break into characteristic classes, perhaps via representationa
of SL(2,Z)$_m$.

It would be interesting to find a differential operator with solutions
that generate the maps $M$ from $E$ to $E_C$, or a recursive construction
of the coefficients pertaining to the maps.  This is possibly obtained via 
topological properties of the manifolds that describe the solution space, 
as found in \cite{ChalmersPoly}.

The counting of the rational points on the elliptic and hyperelliptic surfaces is an
outstanding problem pertinent to various areas of algebra and geometry.
The approach here is more general than finding the cardinality, i.e.
infinite of finite, of the rational points.  This means that the maps
would find all sets of points $x=s/t$ for $p/q$ at given values of $t$
and $p/q$, as opposed to whether the number of solutions is infinite.

The three functions via the torus map are generated via the polynomials,

\bqr
B_C(B,N,D) = \sum_{i,j,k} \beta^{B_C}_{i,j,k} B^i N^i D^i
\fqr
\bqr
N_C(B,N,D) = \sum_{i,j,k} \beta^{N_C}_{i,j,k} B^i N^i D^i
\fqr
\bqr
D_C(B,N,D) = \sum_{i,j,k} \beta^{D_C}_{i,j,k} B^i N^i D^i \ ,
\label{3manifolds}
\fqr
in which the coefficients are rational numbers. The
map is polynomial in nature because the numbers $B_C$, $N_C$ and
$D_C$ are integral.  These maps also define three manifolds via
the polynomial $y^2=B_C$, etc...  These special manifolds define
the counting of the rational solutions to the polynomial equations.
The definitions of the elliptic parameters requires a specification
of the base, in general away from base $2$.  So there is an integral
ambiguity parameterized by $Z$ of the maps.

Due to the three-dimensional interpretation of the function describing
the map between the two torii, it is natural to inquire as to the form
of the moduli space parameterizing the equivalent geometries; meaning
which polynomials $P(z)=p/q$ have the same zero set solutions in the
rational numbers (either in an entire context or partially).  This
is not examined in the text here.

Due to the fact that all of the number solution counting is obtained
from the torus map, encoded in the three functions in \rf{3manifolds}.
It is possible that the numbers $\beta_{i,j,k}$ have some special
number theoretic significance, either recursive or differential
geometric (mentioned that this is possibly obtained via topological 
properties of the associated manifold in \cite{ChalmersPoly}, with related 
background in \cite{ChalmersGeoDiff}), and that 
the homology (i.e. the intersection matrix) of the 3-manifolds defined by 
\rf{3manifolds} is also significant.  This significance should shed
light on a proof of the coefficients and also on the geometric
properties of the distributions of the countings.

The maps are not generated here, however, there are constraints imposed
on their construction.  The solution to the maps is expected to 
be transcendental, and possibly stochastic, due to the parameterizations 
of the relevant numbers; there may be further parameterizations which 
are useful for determinations of the mappings.  Uniformization of the 
curves is interesting.

\vfill\break

\end{document}